\documentclass[floatfix,aps,a4paper,prd,twocolumn,nofootinbib,preprintnumbers]{revtex4-1} 

\usepackage{epsfig}
\usepackage{amsmath}
\usepackage{amssymb}
\usepackage{amsfonts}
\usepackage{graphicx,subfigure}
\usepackage{color}
\usepackage{geometry}
\usepackage{rotating}
\usepackage{psfrag}
\usepackage{bm}
\usepackage{bbm}
\usepackage{multirow}

\newbox\charbox
\newbox\slabox
\def\s#1{{      
 \setbox\charbox=\hbox{$#1$}
 \setbox\slabox=\hbox{$/$}
 \dimen\charbox=\ht\slabox
 \advance\dimen\charbox by -\dp\slabox
 \advance\dimen\charbox by -\ht\charbox
 \advance\dimen\charbox by \dp\charbox
 \divide\dimen\charbox by 2
 \raise-\dimen\charbox\hbox to \wd\charbox{\hss/\hss}
 \llap{$#1$}
}}

\graphicspath{%
{./}%
{figs/}%
}

\newcommand{\newc}{\newcommand}
\newc{\wt}{\widetilde}
\newc{\cL}{{\cal L}}
\newc{\cM}{{\cal M}}
\newc{\ra}{\rightarrow}
\newc{\eps}{\epsilon}
\newc{\bino}{\widetilde{\cal B}}
\newc{\wino}{\widetilde{\cal W}}
\newc{\gluino}{\widetilde{\cal G}}
\newc{\half}{\frac{1}{2}}
\newc{\third}{\frac{1}{3}}
\newc{\fourth}{\frac{1}{4}}
\newc{\eighth}{\frac{1}{8}}
\newc{\gev}{\mbox{~GeV}}
\newc{\lra}{\leftrightarrow}
\newc{\Dslash}{\not\!\! D}
\newc{\sg}{{\cal G}}
\newc{\ovl}{\overline}
\newc{\ok}{$\surd$}
\newc{\etal}{{\it et al.}\ }
\newc{\Hbar}{{\bar H}}
\newc{\hhbar}{{\overline h}}
\newc{\Ubar}{{\bar U}}
\newc{\Dbar}{{\bar D}}
\newc{\Ebar}{{\bar E}}
\newc{\eg}{{\it e.g.}\ }
\newc{\ie}{{\it i.e.}\ }
\newc{\nonum}{\nonumber}
\newc{\kap}{\kappa}
\newc{\Dt}{\frac{d}{dt}}
\newc{\rpv}{{\mbox{${\not\!\!R_p}$}}}
\newc{\bpv}{$\not\!\!B_p$}
\newc{\mpl}{$M_{Pl}$\ }
\newc{\mx}{$M_X$\ }
\newc{\mgut}{M_\mathrm{GUT}}
\newc{\tev}{\mbox{~TeV}}
\newc{\sect}[1]{\ref{sec:#1}}
\newc{\nonr}{\nonum}
\newc{\vev}[1]{\langle{#1}\rangle}
\newc{\eq}[1]{(\ref{eq:#1})}
\newc{\eqs}[2]{(\ref{eq:#1},\ref{eq:#2})}
\newc{\lab}[1]{\label{eq:#1}}
\newc{\Lam}{{\bf \Lambda}}
\newc{\ltau}{\lambda_\tau}
\newc{\lt}{\lambda_t}
\newc{\lb}{\lambda_b}
\newc{\lae}{{\Lam}_E}
\newc{\lad}{{\Lam}_D}
\newc{\lau}{{\Lam}_U}
\newc{\lame}[1]{{\Lam}_{E^{#1}}}
\newc{\lamhe}[1]{{\h}_{E^{#1}}}
\newc{\lamhed}[1]{{\h}_{E^{#1}}^\dagger}
\newc{\lamhd}[1]{{\h}_{D^{#1}}}
\newc{\lamhdd}[1]{{\h}_{D^{#1}}^\dagger}
\newc{\lamhu}[1]{{\h}_{U^{#1}}}
\newc{\lamhud}[1]{{\h}_{U^{#1}}^\dagger}
\newc{\lamd}[1]{{\Lam}_{D^{#1}}}
\newc{\lamu}[1]{{\Lam}_{U^{#1}}}
\newc{\lamet}[1]{{\Lam}_{E^{#1}}^T}
\newc{\lamdt}[1]{{\Lam}_{D^{#1}}^T}
\newc{\lamut}[1]{{\Lam}_{U^{#1}}^T}
\newc{\lames}[1]{{\Lam}_{E^{#1}}^*}
\newc{\lamds}[1]{{\Lam}_{D^{#1}}^*}
\newc{\lamus}[1]{{\Lam}_{U^{#1}}^*}
\newc{\lamed}[1]{{\Lam}_{E^{#1}}^\dagg}
\newc{\lamdd}[1]{{\Lam}_{D^{#1}}^\dagg}
\newc{\lamud}[1]{{\Lam}_{U^{#1}}^\dagg}
\newc{\lam}{{\bf \lambda}}
\newc{\lamp}{{\bf \lambda}^{\prime}}
\newc{\lampp}{{\bf \lambda}^{\prime\prime}}
\newc{\Y}{{\bf Y}}
\newc{\h}{{\bf h}}
\newc{\meee}{{{\rm {\bf  m}}_e}}
\newc{\mdee}{{{\rm {\bf  m}}_d}}
\newc{\myew}{{{\rm {\bf m}}_u}}
\newc{\ye}{{\Y}_E}
\newc{\he}{{\h}_E}
\newc{\hed}{{\h}_E^\dagger}
\newc{\yd}{{\Y}_D}
\newc{\hd}{{\h}_D}
\newc{\hdd}{{\h}_D^\dagger}
\newc{\yu}{{\Y}_U}
\newc{\hu}{{\h}_U}
\newc{\hud}{{\h}_U^\dagger}
\newc{\yes}{{\Y}_E^*}
\newc{\yds}{{\Y}_D^*}
\newc{\yus}{{\Y}_U^*}
\newc{\yet}{{\Y}_E^T}
\newc{\ydt}{{\Y}_D^T}
\newc{\yut}{{\Y}_U^T}
\newc{\yed}{{\Y}_E^\dagg}
\newc{\ydd}{{\Y}_D^\dagg}
\newc{\yud}{{\Y}_U^\dagg}
\newc{\dagg}{\dagger}
\newc{\lp}{\left(}
\newc{\rp}{\right)}
\newc{\inv}{\frac{1}{16\pi^2}}
\newc{\invsq}{\frac{1}{(16\pi^2)^2}}
\newc{\ggam}[2]{\gamma_{#2}^{#1}}
\newc{\yukgam}[2]{\inv \gamma_{#1}^{(1){#2}}+\invsq\gamma_{{#1}}^{(2){#2}}}
\newc{\susyunif}{ohman,nirpaul,marcelacarlos,susyunif}
\newc{\lsim}{\stackrel{<}{\sim}}
\newc{\gsim}{\stackrel{>}{\sim}}
\newc{\Tr}{{~\rm Tr}}
\newc{\me}{{(\bf m_{\tilde{E}}}^2)}
\newc{\mh}[1]{m_{H_{#1}}^2}
\newc{\ml}{{\bf m_{\tilde{L}}}^2}
\newc{\md}{{(\bf m_{\tilde{D}}}^2)}
\newc{\mup}{{(\bf m_{\tilde{U}}}^2)}
\newc{\mq}{{(\bf m_{\tilde{Q}}}^2)}
\newc{\mlh}[1]{{\bf m}_{ \tilde{L}_{#1} H_1}^2}
\newc{\mhl}[1]{{\bf m}_{ H_d \tilde{L}_{#1}}^2}
\newc{\del}{\partial}
\newc{\beq}{\begin{equation}}
\newc{\eeq}{\end{equation}}
\newc{\barr}{\begin{eqnarray}}
\newc{\earr}{\end{eqnarray}}
\newc{\dspl}{\displaystyle}
\newc{\phmin}{\phantom{-}}
\newc{\stau}{{\tilde\tau}}
\newc{\mnu}{$m_{\nu}$ }
\newc{\AO}{$A_0$ }
\newc{\vd}{$v_d$ }
\newc{\MGUT}{$M_\mathrm{GUT}$}
\newc{\mzero}{M_0}
\newc{\mhalf}{{M_{1/2}}}
\newc{\tanb}{\tan\beta}
\newc{\azero}{A_0}
\newc{\sgnmu}{\textrm{sgn}(\mu)}
\def\slashchar#1{\setbox0=\hbox{$#1$}     		
   \dimen0=\wd0                                 	
   \setbox1=\hbox{/} \dimen1=\wd1               	
   \ifdim\dimen0>\dimen1                        	
      \rlap{\hbox to \dimen0{\hfil/\hfil}}      	
      #1                                        	
   \else                                        	
      \rlap{\hbox to \dimen1{\hfil$#1$\hfil}}   	
      /                                         	
   \fi}
\newc{\cf}{\textit{cf.}~}     
\newc{\dzero}{D\O}
\newc{\etmiss}{\slashchar{E}_T}
\newc{\Psix}{{\mathrm{P}_{\!6}}}
\newc{\nPsix}{{\not\!\Psix}}
\newc{\nPsixU}{{\not{\mathrm P}_{\!6}}}
\newc{\Bthree}{{\mathrm{B}_{\,\!3}}}
\newc{\mj}{m_{\tilde{N}_j}}
\newc{\mk}{m_{\tilde{N}_k}}
\newc{\slepton}{\tilde \ell}

\makeatletter

\newcommand{\Rmnum}[1]{\expandafter\@slowromancap\romannumeral #1@}
\makeatother

\newc{\squark}{\tilde{q}}
\newc{\ssup}{\tilde{u}}
\newc{\ssdown}{\tilde{d}}
\newc{\ssstrange}{\tilde{s}}
\newc{\sscharm}{\tilde{c}}
\newc{\sstop}{\tilde{t}}
\newc{\ssbottom}{\tilde{b}}
\newc{\sse}{\tilde{e}}
\newc{\ssmu}{\tilde{\mu}}
\newc{\sstau}{\tilde{\tau}}
\newc{\ssnue}{\tilde{\nu}_{e}}
\newc{\ssnumu}{{\tilde{\nu}_{\mu}}}
\newc{\ssnutau}{{\tilde{\nu}_{\tau}}}
\newc{\ssbnue}{\bar{\tilde{\nu}}_{e}}
\newc{\ssbnumu}{\bar{\tilde{\nu}}_{\mu}}
\newc{\ssbnutau}{\bar{\tilde{\nu}}_{\tau}}
\newc{\neut}{{\tilde{\chi}}^0}
\newc{\charge}{\tilde{\chi}}
\newc{\glu}{\tilde{g}}
\newc{\Higgs}{H^0}

\newc{\nue}{\nu_e}
\newc{\numu}{\nu_{\mu}}
\newc{\nutau}{\nu_{\tau}}
\newc{\bnue}{\bar{\nu}_e}
\newc{\bnumu}{\bar{\nu}_{\mu}}
\newc{\bnutau}{\bar{\nu}_{\tau}}
\definecolor{Gray}{gray}{0.5}

\begin{document}


\title{Constraining Selectron LSP Scenarios with Tevatron Trilepton Searches}
\author{H.~K.~Dreiner}
\email[]{dreiner@th.physik.uni-bonn.de}
\affiliation{Bethe Center for Theoretical Physics and Physikalisches 
Institut, Universit\"at Bonn, Bonn, Germany}

\author{S.~Grab}
\email[]{sgrab@scipp.ucsc.edu}
\affiliation{SCIPP, University of California Santa Cruz, Santa Cruz, 
CA 95064, USA}

\author{T.~Stefaniak}
\email[]{tim@th.physik.uni-bonn.de}
\affiliation{Bethe Center for Theoretical Physics and Physikalisches 
Institut, Universit\"at Bonn, Bonn, Germany and \Rmnum{2}. Physikalisches Institut, Universit\"at G\"ottingen, G\"ottingen, Germany}

\begin{abstract}
The Tevatron collaborations have searched for associated production of
charginos and neutralinos via trilepton final states.  No events above
the Standard Model prediction were observed.  We employ these results
to put stringent bounds on $R$-parity violating models with a
right-handed scalar electron as the lightest supersymmetric
particle. We work in the framework of lepton number violating minimal
supergravity. We find that within these models the complete parameter
space consistent with the anomalous magnetic moment of the muon can be
excluded at $90\%$ confidence level.  We also give prospects for
Tevatron trilepton searches assuming an integrated luminosity of 10
fb$^{-1}$. We find that Tevatron will be able to test selectron LSP
masses up to 170~GeV.
\end{abstract}

\preprint{BONN--TH--2011--06, SCIPP 11/02}

\maketitle

\section{Introduction}

The LHC has been running for over a year and first searches for
supersymmetry \cite{Nilles:1983ge,Haber:1984rc} have been published~\cite{Khachatryan:2011tk,Collaboration:2011hh,Collaboration:2011qk,
Collaboration:2011wc,Collaboration:2011bz}. In
order to know what can possibly be expected at the LHC with present
and forthcoming data, it is important to know the bounds implied by
existing Tevatron searches
\cite{Aaltonen:2008pv,Abazov:2009zi,Forrest:2009gm}. It is our purpose
here to investigate the bounds from Tevatron trilepton searches
\cite{Baer:1986dv,Baer:1986vf,Barger:1998hp,Matchev:1999nb,Baer:1999bq,Dedes:2002zx}
on a specific supersymmetric scenario.

When extending the Standard Model (SM) of particle physics to include
supersymmetry and implementing the minimal particle content, the
supersymmetric Standard Model has more than 200 new parameters. Most
of these arise from the supersymmetry breaking sector~\cite{Nilles:1983ge,Haber:1984rc}. In order to be able to perform 
phenomenological studies, usually simpler models are considered. We
focus here on the baryon triality (B$_3$) mSUGRA model~\cite{Allanach:2003eb,Allanach:2006st}, where B$_3$ is theoretically
well motivated as an anomaly--free discrete gauge symmetry~\cite{Dreiner:2007vp}. It has only 6 new parameters at the grand
unification (GUT) scale ($M_{\rm
GUT}=\mathcal{O}(10^{16}\,\mathrm{GeV})$)
\begin{align}
\mzero,\,\mhalf,\,\azero,\,\tanb,\,\sgnmu,\,\mathbf{\Lambda}. \label{Eqn:B3mSUGRA}
\end{align}
Here, $\mzero$, $\mhalf$ and $\azero$ are the universal scalar mass,
the universal gaugino mass and the universal trilinear scalar
coupling, respectively. $\tanb$ denotes the ratio of the two Higgs
vacuum expectation values (vevs), and $\sgnmu$ fixes the sign of the
bilinear Higgs mass parameter $\mu$. $\mathbf {\Lambda}$ is a
lepton--number and $R$--parity violating parameter described
below.

In $\Bthree$ mSUGRA, the superpotential is extended by the lepton
number violating (LNV) terms \cite{Dreiner:1997uz},
\begin{align}
W_\mathrm{LNV} = \frac{1}{2} \lam_{ijk} L_i L_j \bar E_k + 
\lamp_{ijk} L_i Q_j \bar D_k + \kappa_i L_i H_2, 
\label{Eqn:WB3}
\end{align}
which are absent in the minimal supersymmetric standard model (MSSM).
Here, $L_i,\, Q_i,\,H_{2},\,\bar E_i$ and $\bar D_i$ are the standard
MSSM chiral superfields. $i,j,k$ are generation indices.  $\lam_{ijk}$
and $\lamp_{ijk}$ are dimensionless couplings. The $\kappa_i$ are
dimensionful parameters, which vanish in $\Bthree$ mSUGRA at
$M_\mathrm{GUT}$ due to a redefinition of the lepton and Higgs
superfields~\cite{Allanach:2003eb}. They are generated at lower scales
via the renormalization group equations (RGEs), leading to interesting
phenomenological consequences for neutrino masses
\cite{Allanach:2007qc,Dreiner:2010ye}.

In the $\Bthree$ mSUGRA model, we assume that exactly one of the
thirty-six dimensionless couplings in Eq.~(\ref{Eqn:WB3}) is non-zero
and positive at the GUT scale. The parameter $\mathbf{\Lambda}$ in
Eq.~(\ref{Eqn:B3mSUGRA}) refers to this choice, \ie
\begin{align}
\mathbf{\Lambda} \in \{ \lam_{ijk}, \, \lamp_{ijk}\}, \qquad i,j,k = 1,2,3.
\end{align}
Given one coupling at $M_{\mathrm{GUT}}$, other couplings that violate
only the same lepton number are generated at the weak scale, $M_Z$,
through the RGEs~\cite{Allanach:2003eb,Dreiner:2008rv,Allanach:1999mh,deCarlos:1996du}.

\begin{table}
\centering
\setlength{\tabcolsep}{0.0pc}
\begin{tabular*}{0.48\textwidth}{@{\extracolsep{\fill}}lcc}
\hline\hline
$L_iL_j\bar E_k$ & LSP candidate & $2\sigma$ bound \\
\hline
$\lam_{121}$, $\lam_{131}$		& $\sse_R$	&	$0.020\times (M_{\sse_R}/100\gev)$	\\
 $\lam_{231}$& $\sse_R$	&	$0.033\times (M_{\sse_R}/100\gev)$	\\
$\lam_{132}$			& $\ssmu_R$	& 	
$0.020\times (M_{\ssmu_R}/100\gev)$	\\ 
\hline\hline
\end{tabular*}
\caption{List of $L_iL_j\bar E_k$ couplings (first column) needed to 
generate a $\sse_R$- or $\ssmu_R$-LSP (second column). The third
column gives the most recent experimental bounds [$95\%$ 
confidence level (C.L.)], taken from Ref.~\cite{Kao:2009fg}. The
bounds apply at $M_{\rm GUT}$. }
\label{Tab:lamcouplings}
\end{table}

An important feature of B$_3$ mSUGRA models is that the
lightest supersymmetric particle (LSP) is no longer stable.
It is therefore not restricted to be electrically and color neutral~\cite{Ellis:1983ew}. Any supersymmetric particle can be the LSP.
LNV interactions can significantly alter the RGE running of the
sparticle masses such that we obtain new candidates for the LSP beyond
the lightest neutralino, $\tilde{\chi}_1^0$, and lightest stau,
$\tilde{\tau}_1$~\cite{Bernhardt:2008jz,Dreiner:2008ca}. We recently
showed in detail in Ref.~\cite{Dreiner:2011wm} that the interplay of a
large magnitude of (negative) $\azero$ with a $L_i L_j\bar E_k$
coupling $\mathbf{\Lambda} \gtrsim \mathcal{O}(10^{-2})$ can lead to a
right-handed slepton LSP, $\tilde{\ell}_R$, of the first or second
generation, \textit{i.e.} to a selectron, $\tilde{e}_R$, or smuon,
$\tilde{\mu}_R$, LSP. The respective $L_iL_j\bar E_k$ couplings are
given in Table~\ref{Tab:lamcouplings} with their most recent $2\sigma$
upper bounds~\cite{Kao:2009fg} at $M_{\mathrm{GUT}}$.

We also showed in Ref.~\cite{Dreiner:2011wm} that our $\tilde{\ell}_R$
LSP scenarios naturally lead to multi-lepton final states at hadron
colliders. We found that the LHC can test large regions of the
$\tilde{\ell}_R$ LSP parameter space even with first data. These
promising results have motivated us to investigate the present bounds
on our model from Tevatron trilepton searches~\cite{Abazov:2009zi,Aaltonen:2008pv,Forrest:2009gm}: this is the topic
of this paper. To be specific, we will concentrate on selectron LSP
scenarios, where $\lam_{231}$ is the dominant $R$-parity violating
coupling at $M_{\rm GUT}$. Due to the weaker experimental bound on
$\lam_{231}$, \cf Table~\ref{Tab:lamcouplings}, we can obtain a
lighter sparticle mass spectrum resulting in larger cross sections for
sparticle pair production at the Tevatron.

We find that the Tevatron rules out the $\sse_R$ LSP parameter
space within $\text{B}_3$ mSUGRA, which is consistent with the
anomalous magnetic moment of the muon within 2 standard deviations\footnote{The respective $\text{B}_3$ mSUGRA parameter space with a $\ssmu_R$ LSP is already ruled out by the stronger bound on $\lambda_{132}$, \cf Table~\ref{Tab:lamcouplings}.}. 
One should thus also consider going beyond B$_3$ mSUGRA. We extra\-polate
the existing Tevatron analysis to an integrated luminosity of 10
fb$^{-1}$ and find that more statistics can highly improve the
sensitivity for heavier models. We therefore hope to encourage the
Tevatron collaborations to search for our models in their
upcoming trilepton supersymmetry (SUSY) searches.

This paper is organized as follows. In Sec.~\ref{Sec:selectron_LSP},
we review the $\sse_R$ LSP parameter space relevant for our analysis
and develop two benchmark scenarios for the Tevatron. We then apply
in Sec.~\ref{Sec:Tevatron_constraints} the most recent D\O~trilepton
search~\cite{Abazov:2009zi} to the benchmark points and show the
$\sse_R$ LSP parameter space excluded by the Tevatron. In
Sec.~\ref{Sec:prospects}, we give prospects for future Tevatron
analyses. We conclude in Sec.~\ref{Sec:summary}.
Appendix~\ref{App:benchmarks} presents the sparticle masses and
branching ratios for our benchmark models.

\section{The Selectron LSP in $R$-parity Violating mSUGRA}
\label{Sec:selectron_LSP}

\subsection{Selectron LSP Parameter Regions}
\label{Sect:parameterspace}

\begin{figure}
	\centering \includegraphics[scale=1.0]{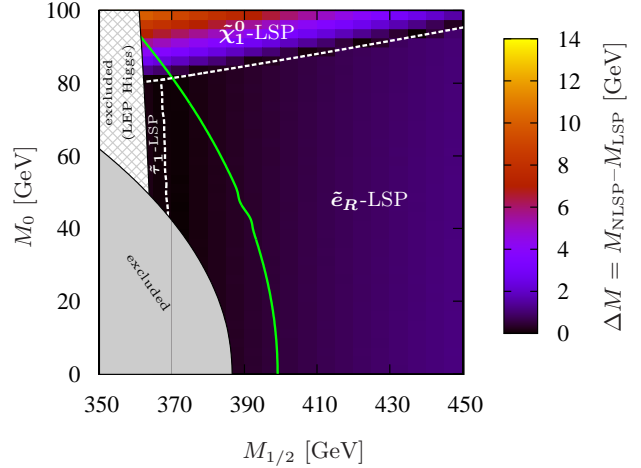}
	\caption{Mass difference, $\Delta M$, between the next--to LSP
	(NLSP) and LSP in the $\mhalf-\mzero$ plane. The other
	$\Bthree$ mSUGRA parameters are $\azero=-1250\gev$, $\tanb=5$,
	$\sgnmu = +$ and $\lam_{231}|_\mathrm{GUT} = 0.045$. The LSP
	candidate regions are shown, bordered by the white dotted
	lines. The solid gray region on the bottom left is excluded
	due to the bound on $\lambda_{231}$, \cf
	Table~\ref{Tab:lamcouplings}, and the patterned region in the
	top left is excluded by Higgs searches at LEP. The
	green contour line indicates the SUSY contribution to the
	anomalous magnetic moment of the muon, $\delta
	a_\mu^\mathrm{SUSY}$. Models to the left lie within the
	$2\sigma$ window for $\delta a_\mu^\mathrm{SUSY}$, \cf
	Eq.~\eqref{Eqn:amu}.}
\label{Fig:m0m12}
\end{figure}

The phenomenology and the typical parameter space of $\Bthree$ mSUGRA
models with a $\sse_R$ or $\ssmu_R$ LSP was discussed in detail in
Ref.~\cite{Dreiner:2011wm}.  We review here the parameter space
relevant for this work. In Fig.~\ref{Fig:m0m12} we show a typical
$\Bthree$ mSUGRA parameter region with a $\sse_R$ LSP in the
$\mhalf-\mzero$ plane. We have chosen a fairly large negative value of
$\azero=-1250\gev$, in order to enhance the (negative) effect of
$\lam_{231}$ on the RGE running of the $\sse_R$ mass. The other
parameters are $\tanb=5$, $\sgnmu=+$ as well as $\lam_{231} = 0.045$
at the GUT scale. We can identify a $\sse_R$, a $\sstau_1$ and a
$\neut_1$ LSP region. The solid gray region at low values of $\mhalf$
and $\mzero$ is excluded by the bound on the LNV coupling, \cf
Table~\ref{Tab:lamcouplings}. The green contour line indicates the
lower value of the $2\sigma$ window (using pion spectral
functions from $e^+e^-$ data\footnote{Note that the SM
prediction of $a_\mu$ is consistent with observations if one uses
spectral functions from $\tau$ data \cite{MALAESCU:2010ne}.}) of the
SUSY contribution to the anomalous magnetic moment of the
muon~\cite{Stockinger:2007pe},
\begin{align}
11.9 \times 10^{-10} < \delta a_\mu^\mathrm{SUSY} < 47.1 \times 10^{-10},
\label{Eqn:amu}
\end{align}
\ie parameter points left of the green line lie within the $2\sigma$ 
window and thus give a significant SUSY contribution to $a_\mu$.
Furthermore, the entire displayed region fulfills the $2\sigma$
constraints on the branching ratios of the decay $b \to s
\gamma$~\cite{TheHeavyFlavorAveragingGroup:2010qj},
\begin{align}
3.03 \times 10^{-4} < \mathcal{B}(b\to s \gamma) < 4.07 \times
10^{-4},
\end{align}
and the $95\%~\mathrm{C.L.}$ upper limit on the
flavor-changing-neutral-current (FCNC) decay $B_s^0\to\mu^+\mu^-
$~\cite{Morello:2009wp},
\begin{align}
\mathcal{B}(B_s^0 \to \mu^+\mu^-) < 3.6 \times 10^{-8}.
\end{align}
We also consider the bounds from Higgs searches at LEP on the light
Higgs mass~\cite{Schael:2006cr}. We employ {\tt 
FeynHiggs2.7.4}~\cite{FeynHiggs} for the calculation of the Higgs
mass, as well as its production and decay properties.  The excluded
supersymmetric parameter space provided by {\tt
HiggsBounds2.1.0}~\cite{HiggsBounds} is indicated as the patterned
region in Fig.~\ref{Fig:m0m12}.  We use {\tt
SOFTSUSY3.0.13}~\cite{Allanach:2001kg,Allanach:2009bv} to calculate
the SUSY spectrum and the Higgs mass parameters, and employ {\tt
micrOMEGAs2.2}~\cite{Belanger:2008sj} to calculate $\mathcal{B}(b\to
s\gamma)$, $\mathcal{B}(B_s^0\to \mu^+\mu^-)$ and $\delta
a_\mu^\mathrm{SUSY}$.

In this study, we focus on light $\sse_R$ LSP models with an LSP mass
$M_\mathrm{LSP} \lesssim 200\gev$. A general feature of these
scenarios is a near mass degeneracy\footnote{A larger mass difference
between $\sse_R$ and $\stau_1$ can be obtained by increasing
$\mathbf{\Lambda}$ and/or $\mhalf$. However, larger values of
$\mathbf{\Lambda}$ translate into the need for larger $\sse_R$
masses, \cf Table~\ref{Tab:lamcouplings}. Thus, increasing either
$\mathbf{\Lambda}$ or $\mhalf$ leads to heavier scenarios, which we
do not consider here.} of the $\sse_R$ LSP with the lightest stau,
$\stau_1$. Thus, a large portion of the $\sse_R$ LSP region exhibits
a $\stau_1$ NLSP. However, close to the $\neut_1$ LSP region at
larger values of $\mzero$, we have $\sse_R$ LSP scenarios with
$M_{\sse_R} \lesssim M_{\neut_1} \lesssim M_{\stau_1}$, \ie a
$\neut_1$ NLSP.
	
\subsection{Benchmark Scenarios for Tevatron Searches}

\begin{table}
\centering
\setlength{\tabcolsep}{0.0pc}
\begin{tabular*}{0.48\textwidth}{@{\extracolsep{\fill}}lcc}
\toprule
$\Bthree$ mSUGRA parameter & SUSY1	&SUSY2 \\
\colrule
$\mzero$ [GeV]			& $0$	&	$80$	\\
$\mhalf$ [GeV]				& $400$	&	$375$	\\
$\azero$ [GeV]				& $-1250$ & $-1250$	\\
$\tanb$					& $5$	& $5$	\\
$\sgnmu$					& $+$	& $+$	\\
$\lam_{231}|_\mathrm{GUT}$	& $0.045$	&$0.045$\\
\botrule
\end{tabular*}
\caption{$\Bthree$ mSUGRA parameters for the benchmark points
SUSY1 and SUSY2.}
\label{Tab:benchmarks}
\end{table}

In this section, we select two benchmark scenarios which we test
explicitly against the D\O~trilepton analysis described in the next
section.  The $\Bthree$ mSUGRA parameters for the two benchmark
points, denoted SUSY1 and SUSY2, are given in
Table~\ref{Tab:benchmarks}. In both scenarios, the dominant $R$-parity
violating coupling is $\lam_{231}= 0.045$ at $M_{\mathrm{GUT}}$.

The benchmark point SUSY1 represents a wide region of the $\sse_R$ LSP
parameter space, where the mass difference between the $\sse_R$ LSP
and the lightest neutralino, $\neut_1$, is much larger than the mass
difference between the $\sse_R$ LSP and the $\stau_1$ NLSP. The masses
of the $\sse_R$, $\stau_1$ and $\neut_1$ are $139.1\gev$, $139.6\gev$
and $163.3\gev$, respectively. In fact, the next-to-NLSP (NNLSP) is
the right-handed smuon, $\ssmu_R$, with a mass of $156.2\gev$. In
contrast, SUSY2 lies in the boundary region to the $\neut_1$
LSP. Here, all three sparticles $\sse_R$, $\stau_1$ and $\neut_1$ are
nearly degenerate in mass, with masses $151.5\gev$, $151.6\gev$ and
$152.8\gev$, respectively. Due to the low mass difference between
$\neut_1$ and $\sse_R$, we expect the electrons from the decay
$\neut_1\rightarrow \sse_R e$ to be fairly soft, such that many do not
fulfill the preselection criteria
\cite{Dreiner:2011wm}.  Detailed tables containing all sparticle
masses and decay modes for these benchmark models are given in
Appendix~\ref{App:benchmarks}. Both SUSY1 and SUSY2 are chosen such
that they are on the edge of the $2\sigma$ lower value of
$\delta a_\mu^\mathrm{SUSY}$ (green line in Fig.~\ref{Fig:m0m12}).

\section{Constraints from the Tevatron}
\label{Sec:Tevatron_constraints}

At the Tevatron at Fermilab both experiments D\O~\cite{Abazov:2009zi}
and CDF~\cite{Aaltonen:2008pv,Forrest:2009gm} have
searched\footnote{Note, that also other SUSY searches using the
trilepton or (like-sign) dilepton signature have been performed at
D\O~and
CDF~\cite{Abulencia:2007rd,Aaltonen:2007mu,Abazov:2006nw,Abazov:2006ii}. At
the current status, these analyses use at most a dataset corresponding
to $1.1~\mathrm{fb}^{-1}$. Thus, we do not expect these searches to be
more restrictive than those presented here.}  for supersymmetry with
final states containing three charged leptons, using the collected
data of proton-antiproton ($p\overline p$) collisions at a
center-of-mass energy $\sqrt{s}=1.96\tev$, corresponding to an
integrated luminosity of 2.3 $\mathrm{fb}^{-1}$ and 3.2
$\mathrm{fb}^{-1}$, respectively. These analyses were designed for the
measurement of associated production of charginos and
neutralinos~\cite{Beenakker:1999xh} within $R$-parity conserving
mSUGRA, using exclusive trilepton search
channels~\cite{Baer:1986dv,Baer:1986vf,Barger:1998hp,Matchev:1999nb,Baer:1999bq,Dedes:2002zx}. Some
of our lighter models could have led to an observable excess of events
in these searches. Here we investigate quantitatively how these
experimental analyses constrain the $\sse_R$ LSP parameter space.

We follow the D\O~analysis to test the exclusion of $\tilde e_R$ LSP
models. CDF uses a jet veto in the event selection, which is expected
to lead to a reduced signal efficiency for many $\sse_R$ LSP
models\footnote{In order to discriminate the $t\overline t$
background, CDF requires the scalar sum of the jet transverse energies
$\sum E_T(\mbox{jets}) \le 80\gev$ and the number of jets
$N(\mbox{jets})<2$
\cite{Aaltonen:2008pv}. We thus expect SUSY events from sparton
(squark and/or gluino) pair production to be mostly rejected in the
CDF analysis.}.  We therefore concentrate on the D\O~search.
Furthermore, D\O~distinguishes their search channels by the flavor of
the final state leptons. Since, in our models, the final state lepton
flavor multiplicity depends on the choice of the $\mathbf{\Lam}$
coupling, we expect different sensitivities of the D\O~search channels
for different choices of $\mathbf{\Lam}$.

In the next section we describe how we emulate the D\O~analysis and
discuss the major changes to the original analysis. We test the two 
$\sse_R$ LSP benchmark points of Table~\ref{Tab:benchmarks} against 
our analysis in Sec.~\ref{Sect:benchmarktest}. We then
review the results of the D\O~analysis and show the excluded regions
of the $\sse_R$ LSP parameter space.

\subsection{The D\O~Trilepton Analysis}
\label{Sect:dzeroanalysis}

\begin{table*}
\centering
\small
\setlength{\tabcolsep}{0.0pc}
\begin{tabular*}{\textwidth}{@{\extracolsep{\fill}}lccccccc}
\toprule
	&	Selection	&	\multicolumn{2}{c}{$\mu\mu \ell$}	& \multicolumn{2}{c}{$e e \ell$} 	&\multicolumn{2}{c}{$e \mu \ell$}\\
	&			&	low $p_T$	&	high $p_T$&	low $p_T$	&	high $p_T$&	low $p_T$	&	high $p_T$\\	
\colrule
\Rmnum{1}	&	$p_T^{\ell_1}, p_T^{\ell_2}$	&	$>12, >8$	&	$>18, >16$	&	 $>12, >8$ & $>20,>10$	&	$>12,>8$\footnote{$p_T^{\ell_1}$ and $p_T^{\ell_2}$ are electron and muon $p_T$, respectively.}	& $>15,>15$	\\
\colrule
			&	$m_{\ell_1\ell_2}$			&	$\in [20,60]$&	$\in [0,75]$	&	$\in [18,60]$&	$\in[0,75]$&	-	&	-	\\
\Rmnum{2}	&	$\Delta \phi_{\ell_1\ell_2}$	&	$<2.9$		&	$<2.9$	&	$<2.9$	&	$<2.9$	&	-	&	-	\\
\colrule
			&	$\etmiss$				&	$>20$		&	$>20$	&	$>22$	&	$>20$	&	$>20$	&	$>20$\\
	&	{\color{Gray}$\mathrm{Sig}(\etmiss)$}&{\color{Gray}$>8$}&{\color{Gray}$>8$}&{\color{Gray}$>8$}	&{\color{Gray}$>8$}	&{\color{Gray}$>8$}	&{\color{Gray}$>8$}\\
			&	$m_T^{\mathrm{min}}$	&	$>20$		&	$>20$	&	$>20$	&	$>14$	&	$>20$	&	$>15$\\
\Rmnum{3}	&	$H_T$				&	-			&	$<80$	&	-		&	-		&	-		&	-	\\
\colrule
\Rmnum{4}	&	$p_T^{\ell_3}$			&	$>5$			&	$>4$		&	$>4$		&	$>12$	&	$>6$		&	$>6$\\
\colrule
			&	$m_T^{\ell_3}$			&	$>10$		&	$>10$	&	$>10$	&	$>10$	&	$>10$	&	$>8$\\
\Rmnum{5}	&	$m_{\ell_{1,2}\ell_3}$			&	$\not\in[80,110]$	&	-		&	-		&	-		&	$<70$	&	$<70$\\
\colrule
			& {\color{Gray}anti $W$}	&	{\color{Gray}-}	&	{\color{Gray}-}	&	{\color{Gray}tight likelihood\footnote{for $p_T^{\ell_3} < 15\gev$}}	&	{\color{Gray}-}	&	\multicolumn{2}{c}{{\color{Gray}tight likelihood\footnotemark[3]}}\\
			&						&				&			&			&			& \multicolumn{2}{c}{{\color{Gray}hit in 2 inner layers\footnotemark[3]}} \\
			&						&				&			&			&			& \multicolumn{2}{c}{{\color{Gray}very tight muon isolation\footnotemark[4]}} \\
\Rmnum{6}	&						&				&			&			&			& \multicolumn{2}{c}{{\color{Gray}$\sum_{0.05 < \Delta R < 0.4} p_T^{\ell_3} < 1$}}\\
\colrule
			&	$\etmiss \times p_T^{\ell_3}$&	$>200$		&	$>300$	&	$>220$	&	-		&		-	&	-	\\
\Rmnum{7}	&	$p_T^{\mathrm{bal}}$	&	$<4$			&	$<4$		&	$<4$		&	$<4$		&	$<2$		&	$<2$\\
\botrule
\end{tabular*}
\footnotetext{for $m_T^\mu \in [40,\,90]\gev$}
\footnotetext{for $m_T^e \in [40,\,90]\gev$}
\caption{D\O~selection criteria for the $\mu\mu l$, $e e l$ and 
$e\mu l$ analyses for the low-$p_T$ selection and the high-$p_T$
selection, see text and Ref.~\cite{Abazov:2009zi} for further
details. All energies, masses and momenta are in GeV, angles
are in radians. We apply all cuts except the cut on $\mathrm{Sig}(
\etmiss)$ in step \Rmnum{3} and the anti $W$ requirements in step 
\Rmnum{6} (both marked in gray).}
\label{Tab:D0cuts}
\end{table*}

The D\O~search for associated production of charginos and neutralinos
with final states containing three charged leptons is presented in
Ref.~\cite{Abazov:2009zi}.  The analysis is based on $p\overline{p}$
collision data at a center-of-mass energy of $\sqrt{s}=1.96\tev$
corresponding to an integrated luminosity of $2.3\,\mathrm{fb}^{-1}$,
with the exception of the analysis using identified hadronic $\tau$
lepton decays, which is based on $1\, \mathrm{fb}^{-1}$ of data. Four
dedicated trilepton event selections were designed, distinguished by
the lepton content in the final state, \ie we have a $ee \ell$,
$\mu\mu \ell$, $e\mu \ell$ and $\mu \tau$ selection without
specification of the lepton charge. Here the third lepton $\ell$
corresponds to a reconstructed isolated track without using the
D\O~standard lepton identification criteria. The first three channels
are separated into a low-$p_T$ and a high-$p_T$ selection, while the
$\mu \tau$ channel contains a $\mu \tau \ell$ selection and a $\mu
\tau \tau$ selection. In this study, we focus on the $ee \ell$,
$\mu\mu \ell$ and $e\mu \ell$ channels. The $\mu \tau$ selection
turned out to be insensitive to our models.

In our object reconstruction, we use cone isolation criteria for all
leptons, where the cone radius $\Delta R = \sqrt{(\Delta \phi)^2 +
(\Delta \eta)^2}$ is given by the distance in pseudorapidity $\eta$
and azimuthal angle $\phi$. Guided by the D\O~object reconstruction,
an electron (muon\footnote{This isolation criteria corresponds to
\textit{tight} muons in Ref.~\cite{Abazov:2009zi}.}) with
pseudorapidity $|\eta|<3.2$ ($|\eta| < 2.0$) is considered as
isolated, if the scalar sum of the absolute value of the transverse
momenta of all tracks in a cone of $\Delta R=0.4$ does not exceed
$2.5\gev$.  We do not loosen the reconstruction criteria for the third
lepton $\ell$ but demand it to be an isolated electron or muon. Jets
are reconstructed with {\tt
FastJet2.4.1}~\cite{Cacciari:2005hq,Cacciari:web} using the kt
algorithm with $R = 0.4$ and must be within $|\eta| < 2.5$. In our
Monte Carlo (MC) simulation, the missing transverse energy, $\etmiss$,
is calculated as the sum over the transverse momenta of all invisible
particles.

In the following, we describe the general features of the various
steps in the event selection. The details are given in
Table~\ref{Tab:D0cuts} and the specific values should be taken from
this table. For a detailed description of the cuts and their effect on
the SM background, we refer the reader again to
Ref.~\cite{Abazov:2009zi}.

First, each selection requires two identified leptons ($\ell=e,\,\mu$)
with certain minimum transverse momenta $p_T^{\ell_1}$, $p_T^{\ell_2}$
(\Rmnum{1}).  If more then two leptons are identified that satisfy the
$p_T$ criteria, the two leptons with the highest $p_T$ are
considered. Next, constraints on the invariant mass $m_{\ell_1\ell_2}$
and the opening angle $\Delta \phi_{\ell_1\ell_2}$ of the two leptons
are imposed (\Rmnum{2}). This is followed in step (\Rmnum{3}) by
requirements on $\etmiss$, the minimal transverse mass
$m_T^{\mathrm{min}}=\mathrm{min} (m_T^{\ell_1},m_T^{\ell_2})$, where
\begin{align}
m_T^\ell = \sqrt{2p_T^\ell \etmiss [1-\cos\Delta\phi(\ell,\etmiss)]},
\end{align}
 and $H_T$, which is the scalar sum of the $p_T$ of all jets with $p_T
 > 15\gev$. In this step, a further requirement on
 $\mathrm{Sig}(\etmiss)$ is performed in the original D\O~analysis,
 where $\mathrm{Sig}(\etmiss)$ is defined for events with at least one
 jet as
\begin{align}
\mathrm{Sig}(\etmiss) \equiv \frac{\etmiss}{\sqrt{\sum_\mathrm{jets} \sigma^2(E_T^j || \etmiss)}}.
\end{align}
Here, $\sigma^2(E_T^j || \etmiss)$ is the jet energy resolution
projected on the $\not\!\!\vec{p}_T$ direction, \textit{i.e.}
on the direction of the missing transverse momentum
vector.\footnote{Note that in Ref.~\cite{Abazov:2009zi} the
symbol ${\slashchar{\vec{E}}_T}$ is used.}  This cut rejects events
with $\etmiss$ faked by poorly measured jets and thus significantly
reduces the QCD background. In our approach, we do not apply this cut
on $\mathrm{Sig}(\etmiss)$, since we do not have a measure of the jet
energy resolution. However, since the missing transverse energy stems
mostly from the neutrinos coming from the leptonically decaying
$\sse_R$ LSP, the effect of this cut is expected to be small.

In step \Rmnum{4}, we demand an additional third lepton with a softer
$p_T$ requirement. Further cuts on its transverse mass $m_T^{\ell_3}$
and the invariant masses $m_{\ell_{1,2},\ell_3}$ of the third lepton
with one of the preselected leptons are applied (\Rmnum{5}). For some
channels in the original D\O~analysis, step (\Rmnum{6}) includes
further lepton quality requirements using likelihood discriminants in
order to reduce background from $W$ boson production, where the second
lepton is faked by jets or photons. This step is skipped in our
approach, since this requires a more detailed simulation of the
detector, beyond the scope of this work. In the last step (\Rmnum{7})
we apply a cut on the product of the third lepton $p_T$ and $\etmiss$
as well as on the $p_T$ balance
\begin{align}
p_T^{\mathrm{bal}} = \frac{|{\vec{p}_T}^{\;\ell_1}+{\vec{p}_T}^{\;\ell_2}+\vec{\slashchar{p}}_T|}{p_T^{\ell_3}}.
\label{Eqn:ptbalance}
\end{align}

\subsection{D\O~Results and a Test of two Benchmark Scenarios}
\label{Sect:benchmarktest}

\begin{table}
\centering
\renewcommand{\arraystretch}{1.4}\addtolength{\tabcolsep}{0.3cm}
\setlength{\tabcolsep}{0.0pc}
\begin{tabular*}{0.48\textwidth}{@{\extracolsep{\fill}}lcc}
\toprule
Signal cross section (in fb)		&	SUSY1	&	SUSY2	\\
\colrule
$\sigma(p\overline p \to \mbox{sparton pairs})$	& $1.5 \pm 0.1$	&	$8.3 \pm 0.2$	\\	
$\sigma(p\overline p \to \mbox{slepton pairs})$	& $8.5 \pm 0.2 $	&	$6.5 \pm 0.1$	\\
\multirow{2}{*}{\renewcommand{\arraystretch}{0.8}$\sigma\left(p\overline p \to \begin{array}{l} \mbox{gaugino pairs},\\ \mbox{gaugino-sparton}\end{array}\right)$}	& \multirow{2}{*}{$3.8\pm 0.1$}	&	\multirow{2}{*}{$6.1 \pm 0.1$}	\\
								&				&				\\
\colrule
$\sigma(p \overline p \to \mbox{sparticle pairs})$	& $13.8\pm 0.2$	&	$20.9\pm 0.2$\\
\botrule
\end{tabular*}
\caption{Leading-order (LO) signal cross sections for $p\bar p$ 
collisions at a center-of-mass energy of $\sqrt{s}=1.96\tev$ for the
benchmark scenarios SUSY1 and SUSY2. We give the cross section of
sparton (\ie squark and gluino) pair, slepton pair and electroweak
(EW) gaugino pair / EW gaugino-sparton production separately. The last
row gives the total sparticle pair production cross section, which is
the signal process. We employed {\tt HERWIG6.510} to derive the LO cross
sections and for the event simulation. The uncertainties are due to
statistical fluctuations from {\tt HERWIG}.
}
\label{Tab:xsec}
\end{table}

\begin{table*}
\centering
\scriptsize
\setlength{\tabcolsep}{0.0pc}
\begin{tabular*}{\textwidth}{@{\extracolsep{\fill}}lrrrrrrrrrrrr}
\toprule
Selection & \multicolumn{4}{c}{$\mu\mu \ell$} & \multicolumn{4}{c}{$e
e \ell$} &\multicolumn{4}{c}{$e \mu \ell$}\\ & Data & Backgrd. & SUSY1
& SUSY2 &Data & Backgrd. & SUSY1 & SUSY2 &Data & Backgrd. & SUSY1 &
SUSY2\\
\colrule
\Rmnum{1}&	$194006$&	$195557\pm177$	& $6.6$		&	$17.8$	&$235474$&	$232736\pm202$& $19.8$& $11.7$	& $16630$	&$16884\pm75$ &$12.6$	&	$18.2$\\
\Rmnum{2}&	$22766$	&	$26067\pm88$		&	$1.4$	&	$4.2$	&$31365$	&	$27184\pm64$	&	$4.8$&	$2.8$	& 		&		&		&		\\
\Rmnum{3}&	$178$	&	$181\pm6.4$		&	$1.2$	&	$3.9$	&$515$	&	$212\pm12$	& $4.3$	&	$2.6$	&$1191$	&$1177\pm20$&$11.1$&$16.9$\\
\Rmnum{4}&	$7$		&	$2.9\pm0.7$		&	$1.0$	&	$2.8$	&$16$	&	$9.3\pm2.0$	& $3.0$	&	$1.3$	&$22$	&$18.0\pm1.2$	&	$9.9$&$11.0$\\
\Rmnum{5}&	$4$		&	$2.2\pm0.5$		&	$0.6$	&	$2.4$	&$9$	&	$5.9\pm1.7$	&$2.8$	&	$1.3$	&$3$	&$3.5\pm0.5$&	$3.8$	& $3.9$\\
\Rmnum{6}&			&					&			&			&$6$	&	$3.1\pm0.4$	&		&			&$2$	&$1.6\pm0.4$&		&		\\
\Rmnum{7}&	$4$		&	$1.2\pm0.2$		&	$0.5$	&	$1.8$	&$2$	&	$1.8\pm0.2$	&$2.4$	&	$1.2$	&$2$	&$0.8\pm0.2$&	$1.2$&	$1.0$\\	
\botrule
\end{tabular*}
\caption{Numbers of events observed in the data and expected for the 
background (taken from Ref.~\cite{Abazov:2009zi}) and numbers of
signal (SUSY1 and SUSY2, see text) events at various stages of the
analysis for the $\mu\mu \ell$, $e e l$ and $e\mu \ell$ channels and
the low-$p_T$ selection. Each row corresponds to a group of cuts, as
detailed in Table~\ref{Tab:D0cuts}. This is for an integrated
luminosity of $2.3\,\mathrm{fb}^{-1}$.}
\label{Tab:results_low}

\vspace{0.5cm}
\scriptsize
\setlength{\tabcolsep}{0.0pc}
\begin{tabular*}{\textwidth}{@{\extracolsep{\fill}}lrrrrrrrrrrrr}
\toprule
Selection	&	\multicolumn{4}{c}{$\mu\mu \ell$}	& \multicolumn{4}{c}{$e e \ell$} 	&\multicolumn{4}{c}{$e \mu \ell$}\\
		&	Data	&	Backgrd.	& SUSY1	&	SUSY2	&Data	&	Backgrd.	& SUSY1	&	SUSY2	&Data	&	Backgrd.	& SUSY1	&	SUSY2\\			
\colrule
\Rmnum{1}&	$140417$&	$141781\pm120$	& $5.4$	&	$16.2$	&	$171001$&	$170197\pm175$& $19.0$	& $11.1$		& $4617$	&$4709\pm23$ &$10.6$	&	$17.0$\\
\Rmnum{2}&	$10349$	&	$10645\pm51$		&	$1.9$	&	$5.6$	&$8273$	&	$7937\pm39$	&	$6.8$&	$3.9$	& 		&		&		&		\\
\Rmnum{3}&	$173$	&	$176\pm5.7$		&	$1.2$	&	$3.7$	&$244$	&	$264\pm10$	& $6.4$	&	$3.8$	&$727$	&$738\pm11$&$9.8$&$16.0$\\
\Rmnum{4}&	$7$		&	$3.8\pm0.5$		&	$0.9$	&	$2.8$	&$0$	&	$1.5\pm0.3$	& $3.9$	&	$1.8$	&$11$	&$12.7\pm0.9$	&$\mathbf{{8.8}}$&$\mathbf{{10.3}}$\\
\Rmnum{5}&	$4$		&	$2.9\pm0.4$		&	$0.9$	&	$2.8$	&$0$	&	$1.1\pm0.3$	&$3.7$	&	$1.8$	&$2$	&$2.8\pm0.5$&	$3.3$	& $3.6$\\
\Rmnum{6}&			&					&			&			&		&				&		&			&$0$	&$1.0\pm0.2$&		&		\\
\Rmnum{7}&	$4$		&	$2.0\pm0.3$		&	$0.9$	&	$2.4$	&$0$	&	$0.8\pm0.1$	&$\mathbf{{3.4}}$&$1.7$&$0$	&$0.5\pm0.1$&	$0.9$&	$1.0$\\	
\botrule
\end{tabular*}
\caption{
Same as Table~\ref{Tab:results_low}, but for the high-$p_T$ selection.
Signal event yields that exceed the $90\%~\mathrm{C.L.}$ upper
exclusion bound are bold-face.}
\label{Tab:results_high}
\end{table*}

In order to test whether our benchmark models are excluded, we have
generated $2000$ signal events, \textit{i.e.}  the pair production of
all SUSY particles, scaled to an integrated luminosity of
$2.3\,\mathrm{fb}^{-1}$ and apply the simplified D\O~analysis
described above. We employ the Feldman \& Cousins
method~\cite{Feldman:1997qc} to set $90\%~\mathrm{C.L.}$ upper limits
given the number of expected background events and the number of
observed events, both taken from the
D\O~paper~\cite{Abazov:2009zi}. In those cases where the number of
observed events is smaller than the expected background, we take as
the upper limit the $90\%~\mathrm {C.L.}$ \textit{sensitivity},
defined as the average upper limit that would be obtained by an
ensemble of experiments with the expected background and no true
signal, and given in Table~\Rmnum{12} in
Ref.~\cite{Feldman:1997qc}\footnote{For the number of expected
background events $> 15$, we approximate the sensitivity by the
Feldman \& Cousins upper limit for $N_\mathrm{obs}=N_\mathrm{bkg}$.
This is only relevant for the extrapolation to $10\,\mathrm{fb}^{-1}$
in Sec.~\ref{Sec:prospects}.}. We claim a $90\%~\mathrm{C.L.}$
exclusion of the SUSY scenario, if the number of signal events exceeds
this upper confidence limit in any step of the event selection. We do
this comparison separately for all four\footnote{As mentioned before,
the fourth channel including $\tau$ leptons is insensitive. Thus we do
not present the results for this specific channel here.} selection
channels in order to gain some insight into their sensitivity to our
models. Note that in this method, systematic uncertainties are not
taken into account.

For the simulation, we use {\tt
SOFTSUSY3.0.13}~\cite{Allanach:2001kg,Allanach:2009bv} to calculate
the SUSY mass spectra. The {\tt SOFTSUSY} output is fed into {\tt
ISAWIG1.200} and {\tt ISAJET7.64}~\cite{Paige:2003mg} in order to
calculate the decay widths of the SUSY particles including the
relevant $R$-parity violating decays. We have also added some missing
three-body slepton decays to the ISAJET code; see
Ref.~\cite{Dreiner:2011wm} for details. The signal process, \ie
sparticle pair production, was simulated with {\tt
HERWIG6.510}~\cite{Corcella:2000bw,Corcella:2002jc,Moretti:2002eu}.

For the two benchmark models, the leading-order (LO) cross sections of
the following supersymmetric production processes are given in
Table~\ref{Tab:xsec}: sparton (\ie squark and gluino) pair production,
slepton pair production and electroweak (EW) gaugino pair as well as
EW gaugino-sparton production. For the point SUSY1, sparticle
production is dominated by slepton and gaugino production. In contrast,
for SUSY2 the sparton production dominates due to the low mass of the
lightest stop, $M_{\sstop_1} = 304.9\gev$, which decays exclusively to
the lightest chargino and a bottom quark, \cf Table~\ref{Tab:SUSY2}.
As a conservative approach, we only use the LO cross section for the
signal, while the SM background in the
D\O~analysis~\cite{Abazov:2009zi} includes next-to-leading (NLO) and
next-to-NLO corrections. Note, that higher order corrections usually
enhance SUSY particle production at hadron colliders by several
tens of percent \cite{Beenakker:1999xh,NLO}. For the 
calculation of the Feldman \& Cousins confidence limits we employ {\tt
ROOT}~\cite{Brun199781}.

In Table~\ref{Tab:results_low} and Table~\ref{Tab:results_high}, we
review the results from the D\O~analysis and compare them with the
results for the two $\Bthree$ mSUGRA models SUSY1 and SUSY2 for the
low-$p_T$ and the high-$p_T$ selections, respectively.

In all selections, the signal event yield for both benchmark scenarios
is $\lesssim 20$ events after the two lepton requirement (step
\Rmnum{1}) and for an integrated luminosity of $2.3\,\mathrm{fb}
^{-1}$. Thus, the event yields in the first steps
(\Rmnum{1}-\Rmnum{3}) of the analysis are dominated by the
overwhelming SM background. The analysis becomes sensitive to the
signal once we require the third lepton (step
\Rmnum{4} and beyond). Then, the SM background is reduced to
$\mathcal{O} (1 - 20)$ expected events. We now discuss in detail the
D\O~results and the signal event yields of the different selections
after step \Rmnum{4} of the analysis was performed.

In the $\mu\mu \ell$ channel (in both the low-$p_T$ and high-$p_T$
selection), the number of events observed by D\O~is larger
than the number of expected events from the SM background for all
steps beyond cut \Rmnum{4}. Therefore, this channel has intrinsically
a less restrictive impact on the SUSY models. We expect only $\mathcal{O}(1-3)$ signal
events beyond step \Rmnum{4} for both benchmark points. Hence, the
$\mu\mu\ell$ channel cannot exclude the SUSY1 and SUSY2 models.

Note, that SUSY2 yields roughly three times as many events in this
selection as SUSY1. This is due to the enhanced $\sstop_1$ pair
production and their decay to the lightest chargino, as mentioned
above. The chargino decays to the $\ssnumu$ and a muon $21\%$ of the
time, leading to an enhanced number of muons in the signal. However,
in the $\mu\mu\ell$ high-$p_T$ selection, most of the signal events
from sparton-pair production are rejected by the $H_T$ cut in step
\Rmnum{3}. This reduces in particular the SUSY2 event yield, since
here the production of sparton-pairs comprises $40\,\%$ of the signal cross section, \cf Table~\ref{Tab:xsec}.

In the $ee \ell$ channel, the number of observed events is larger
(lower) than the number of expected SM background events in the
low-$p_T$ (high-$p_T$) selection for all steps beyond cut
\Rmnum{4}. For both benchmark scenarios we expect $\mathcal{O}(1-4)$
signal events in these steps of the analysis. Furthermore, the
number of expected signal events for SUSY1 is roughly
two times more than for SUSY2. This is because in SUSY2 the mass
difference between the $\neut_1$ and the $\sse_R$ LSP is small.
Therefore, the electrons from the decay $\neut_1\to\sse_R e$
tend to be soft and fail to pass the $p_T$ criteria in step \Rmnum{1}
of the $eel$ selection. The SUSY1 event yield exceeds the $90\%~
\mathrm{C.L.}$ upper bound in step \Rmnum{7} of the high-$p_T$ $ee\ell$
selection and is therefore excluded by the D\O~trilepton search.

For the low-$p_T$ selection of the $e\mu\ell$ channel, the number of
observed events tends to be larger than the number of expected SM
background events, whereas in the high-$p_T$ selection, the number of
observed events is slightly less. Both the SUSY1 and SUSY2 event yield
exceed the $90\%~\mathrm{C.L.}$ upper limit in step \Rmnum{4} of the
$e\mu\ell$ high-$p_T$ selection. The following steps in the $e\mu\ell$
channel (step \Rmnum{5} and beyond) are not as sensitive to our models
as step \Rmnum{4}, because the cut on the dilepton invariant masses in
step \Rmnum{5} significantly reduces the signal.

In general, the B$_3$ mSUGRA parameter region close to a $\neut
_1$ LSP is more difficult to exclude due to the soft electrons. For 
instance, in step \Rmnum{4} of the $e\mu\ell$ low-$p_T$ selection, the
$90\%$ C.L. upper limit is $13.0$ events, while we expect $11.0$
signal events for SUSY2. However, if we modify the $\mzero$ value of
SUSY2 from $80\gev$ to $75\gev$, \ie we basically change the mass
difference between $\neut_1$ and $\sse_R$ from $1.3\gev$ to $3.7\gev$,
the number of expected signal events increases to $15.2$ events and
the model is excluded.

We conclude, that the D\O~analysis using $2.3~\mathrm{fb}^{-1}$ of
integrated luminosity excludes both benchmark points SUSY1 and
SUSY2 at $90\%~\mathrm{C.L.}$. The most restrictive channels for
$\sse_R$ LSP models with a dominant $\lam_{231}$ coupling are the
$ee\ell$ high-$p_T$ selection (in step \Rmnum{7}) and the $e\mu\ell$
high-$p_T$ selection (in step \Rmnum{4}). In the next section, we
determine the excluded regions of the $\sse_R$ LSP parameter space.

\subsection{Excluded Selectron LSP Parameter Space}
\label{Sect:exclusions}

We now apply the D\O~analysis to a more extensive $\sse_R$ LSP
parameter region. For this, we perform a scan in the $\mhalf-\mzero$
plane with $\mhalf \in [350\gev, 500\gev]$ in steps of $\Delta\mhalf=5
\gev$ and $\mzero \in [0\gev, 120\gev]$ in steps of $\Delta\mzero=2.5
\gev$. We retain $\lam_{231}=0.045$ at $\mgut$. The other $\Bthree$ 
mSUGRA parameter values are $\azero=-1250\gev$, $\tanb=5$ and $\sgnmu
= +$. The scanned $\sse_R$ LSP parameter region was already discussed
in Sec.~\ref{Sect:parameterspace}, \cf Fig.~\ref{Fig:m0m12}. For each
parameter point with a $\sse_R$ LSP, 2000 signal events were generated
and scaled to an integrated luminosity of $2.3~\mathrm{fb}^{-1}$. Then
the $ee\ell$, $\mu\mu\ell$ and $e\mu\ell$ low-$p_T$ and high-$p_T$
event selections were applied\footnote{We did the same for the
$\mu\tau$ selection for an integrated luminosity of
$1.0~\mathrm{fb}^{-1}$.  However, this channel is not capable of
excluding any $\sse_R$ LSP parameter space. Thus, we do not show any
results for the $\mu\tau$ channels.}. At each step of the event
selection, the number of passed events is compared with the
D\O~results as described above. We make this comparison for all event
selection steps once the third lepton is required, \ie for step
\Rmnum{4} and beyond, \cf Table~\ref{Tab:D0cuts}.

In the following figures, the patterned gray regions mark parameter
points with either a neutralino or stau LSP (as indicated in the
figures) which are not considered here. The solid gray region exhibits
a LSP mass of $M_{\sse_R} \lesssim 136\gev$ and is thus excluded by
the bound on the $\lam_{231}$ coupling, \cf 
Table~\ref{Tab:lamcouplings}. The mass of the $\sse_R$ LSP (in$\gev$)
is given by the gray contour lines.

\begin{figure*}
	\centering
	\small
	\subfigure[\,Excluded $\sse_R$ LSP parameter space with the low-$p_T$ selections.\label{Fig:excl_2.3fb_lam231_low}]{
	   	\includegraphics[scale=1.1]{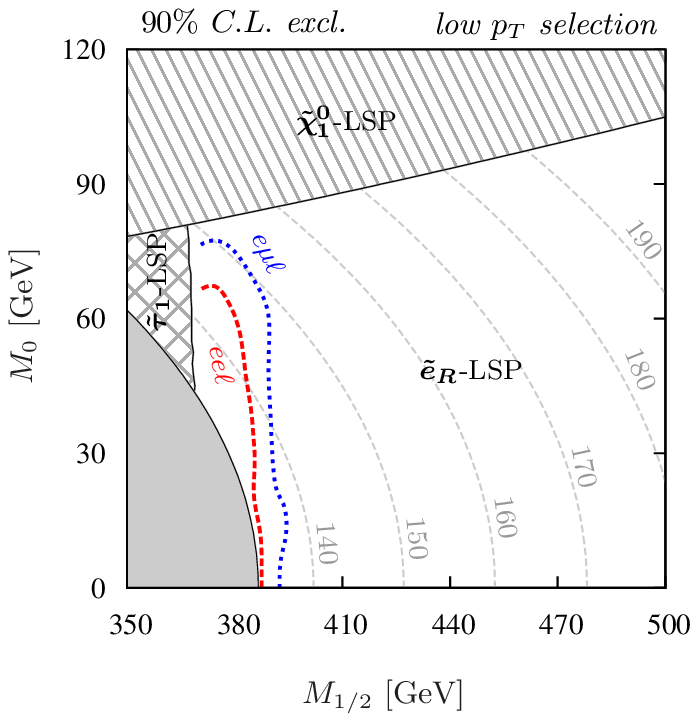}
	}
	\qquad
	\subfigure[\,Excluded $\sse_R$ LSP parameter space with the high-$p_T$ selections.\label{Fig:excl_2.3fb_lam231_high}]{
	   	\includegraphics[scale=1.1]{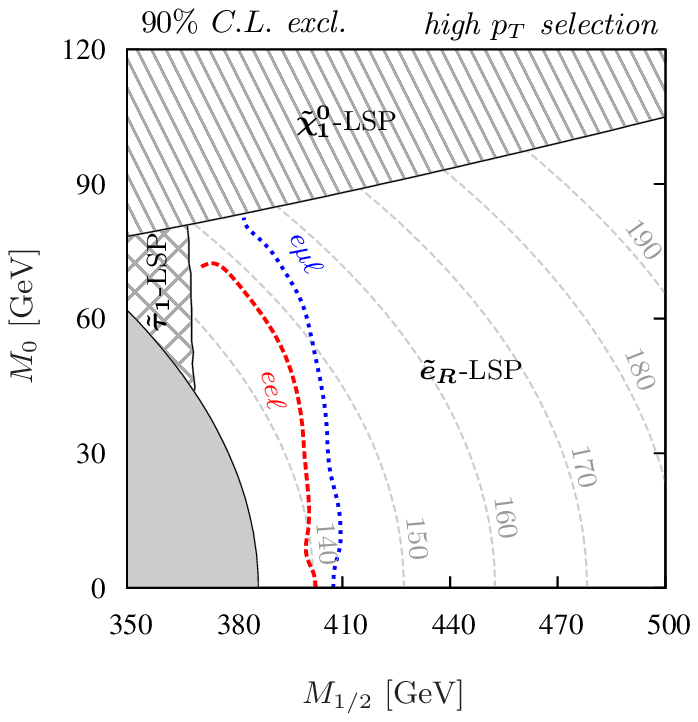}
	}	
\caption{Excluded regions ($90\%~\mathrm{C.L.}$) of the $\sse_R$ LSP 
parameter space by the D\O~trilepton analysis with
$2.3~\mathrm{fb}^{-1}$ of data. We choose $\lam_{231}= 0.045$ as the
dominant LNV coupling at $\mgut$. The other parameters are
$\azero = -1250\gev$, $\tanb=5$ and $\sgnmu =+$. The colored contour
lines give the excluded region by the different channels: In
Fig.~\ref{Fig:excl_2.3fb_lam231_low} they correspond to the $eel$
(red, dashed) and $e\mu l$ (blue, dotted) low-$p_T$ selections, while
in Fig.~\ref{Fig:excl_2.3fb_lam231_high} they are shown for the same
channels in the high-$p_T$ selection. The gray dotted contour lines
give the LSP mass, $M_{\sse_R}$, in GeV, as indicated by the labels.}
\label{Fig:excl_2.3fb_sseR}
\end{figure*}

In Fig.~\ref{Fig:excl_2.3fb_sseR} we give the parameter region that is
excluded at $90\%~\mathrm{C.L.}$ with $2.3~\mathrm{fb}^{-1}$ of
analyzed data. We discuss each channel and $p_T$ selection
separately. Fig.~\ref{Fig:excl_2.3fb_lam231_low}
[Fig.~\ref{Fig:excl_2.3fb_lam231_high}] shows the low-$p_T$
[high-$p_T$] selection of the $ee\ell$ and $e\mu\ell$ channel. The
$\mu\mu \ell$ channel does not exclude any $\sse_R$ LSP parameter
space. 

The LSP decays to $50\%$ to a (hard) muon and a neutrino. Thus, the
$e\mu\ell$ selection is very sensitive to these models and can exclude
$\sse_R$ LSP scenarios with $\sse_R$ masses up to $150\gev$
($155\gev$) and squark masses up to $850\gev$ ($880\gev$) with the
low-$p_T$ (high-$p_T$) selection. The sensitivity decreases for lower
mass differences of the $\neut_1$ and the $\sse_R$ due to the softer
electrons, as can be seen in all displayed channels in
Fig.~\ref{Fig:excl_2.3fb_lam231_low} and
Fig.~\ref{Fig:excl_2.3fb_lam231_high}. Especially the $ee\ell$ channel becomes
insensitive in this boundary region.

\begin{figure*}
	\centering
	\small
	\subfigure[\,$\sse_R$ LSP sensitivity region with the low-$p_T$ selections.\label{Fig:excl_10fb_lam231_low}]{
	   	\includegraphics[scale=1.1]{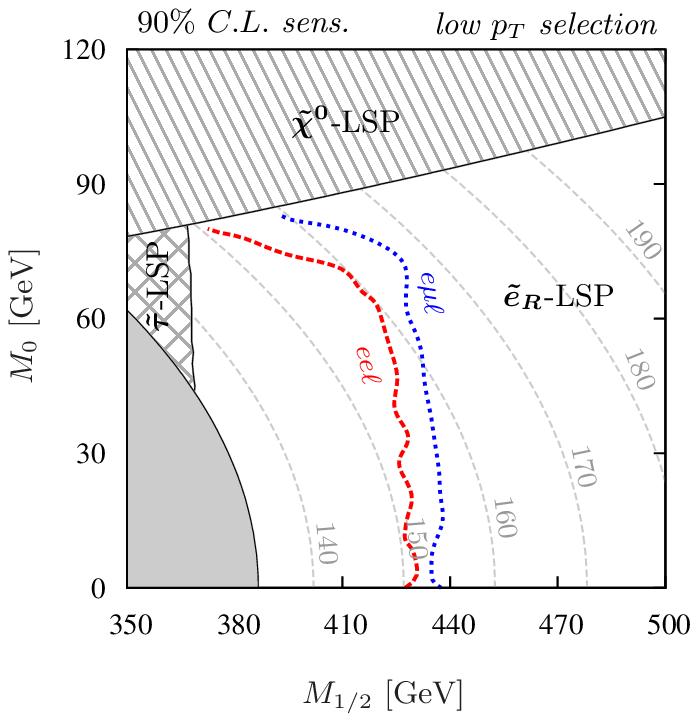}
	}
	\qquad
	\subfigure[\,$\sse_R$ LSP sensitivity region with the high-$p_T$ selections.\label{Fig:excl_10fb_lam231_high}]{
	   	\includegraphics[scale=1.1]{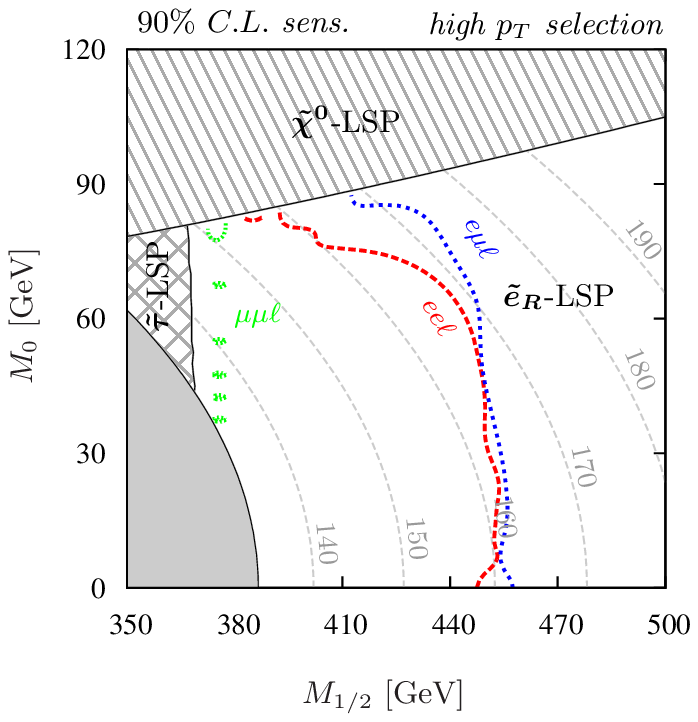}
	} 
\caption{Extrapolated sensitivity regions ($90\%~\mathrm{C.L.}$) of the
$\sse_R$ LSP parameter space for the D\O~trilepton analysis with
future data corresponding to an integrated luminosity of
$10~\mathrm{fb}^{-1}$. The parameter regions are the same as in
Fig.~\ref{Fig:excl_2.3fb_sseR}. The colored contour lines give the
sensitivity of the different channels: In
Fig.~\ref{Fig:excl_10fb_lam231_low} they correspond to the $eel$ (red,
dashed) and $e\mu l$ (blue, dotted) low-$p_T$ selections, while in
Fig.~\ref{Fig:excl_10fb_lam231_high} they are shown for the same
channels in the high-$p_T$ selection. Furthermore, the fine dotted,
green contour line in Fig.~\ref{Fig:excl_10fb_lam231_high} gives the
sensitivity of the $\mu\mu l$ high-$p_T$ selection. The gray dotted
contour lines give the LSP mass, $M_{\sse_R}$, in GeV, as
indicated by the labels.}
\label{Fig:excl_10fb_sseR}
\end{figure*}

Comparing the excluded $\sse_R$ parameter region in
Fig.~\ref{Fig:excl_2.3fb_sseR} with Fig.~\ref{Fig:m0m12}, we conclude
that $\sse_R$ LSP scenarios with a significant contribution to the
anomalous magnetic moment of the muon (region to the left of
the green line in Fig.~\ref{Fig:m0m12}) are excluded at
$90\%~\mathrm{C.L.}$ by the D\O~search with $2.3~\mathrm{fb}^{-1}$ of
analyzed data\footnote{The D\O~search rules out all other
regions of the $\text{B}_3$ mSUGRA parameter space with a $\sse_R$ LSP
consistent with $a_\mu$ (beyond Fig.~\ref{Fig:m0m12}).  For $\tan
\beta \lesssim 4$ these scenarios are ruled out by the LEP Higgs mass
bounds. For $\tan \beta \gtrsim 5.2$ the $\tilde\tau_1$ is the LSP.
$A_0$ is strongly constrained by the requirements presented in
Ref.~\cite{Dreiner:2011wm} and sgn($\mu)=-$ is totally ruled out,
because SUSY will then give a negative contribution to $a_\mu$~\cite{Stockinger:2007pe}. If we go beyond B$_3$ mSUGRA 
then there is still a large parameter region
with a $\tilde e_R$ LSP allowed~\cite{Dreiner:2011wm}.}.

\section{Prospects for Future Tevatron Searches}
\label{Sec:prospects}

Both Tevatron experiments D\O~and CDF acquired $\sim10~\mathrm{fb}^
{-1}$ of data by the end of 2010. Therefore, we extrapolate the
current D\O~results to study the prospects of an exclusion of $\sse_R$
LSP models, using data corresponding to an integrated luminosity of
$10~\mathrm{fb}^{-1}$.

We assume that the events after each selection step in each channel
are observed in the same rate as given by the results with integrated
luminosity of $2.3~\mathrm{fb}^{-1}$, \cf Table~\ref{Tab:results_low}
and Table~\ref{Tab:results_high}. Then, we can extrapolate the data to
the higher integrated luminosity of $10~\mathrm{fb}^{-1}$. By applying
the same method as in the previous section, we determine the
$90\%~\mathrm{C.L.}$ sensitivity region, \ie the supersymmetric
parameter region which would lead to a significant deviation from the
extrapolated data, assuming no discrepancies are observed.

In Fig.~\ref{Fig:excl_10fb_sseR} we present the $\sse_R$ LSP parameter
space, which can potentially be excluded with a future integrated
luminosity of $10~\mathrm{fb}^{-1}$. The parameter space is the same
as in Fig.~\ref{Fig:excl_2.3fb_sseR}. The $90\%~\mathrm{C.L.}$
sensitivity regions for the channels $ee\ell$, $e\mu \ell$ and $\mu\mu
\ell$ are given by the contour lines for the low-$p_T$
(Fig.~\ref{Fig:excl_10fb_lam231_low}) and high-$p_T$
(Fig.~\ref{Fig:excl_10fb_lam231_high}) selection.

The most sensitive channels are the $e\mu\ell$ and $ee\ell$ high-$p_T$
selections, which may exclude scenarios with $\mhalf \lesssim 450\gev$
with future data, assuming no deviation from the SM prediction is
observed. This corresponds to LSP masses $M_{\sse_R}\lesssim(160-170)
\gev$ and squark masses $M_{\squark} \lesssim (900 - 950)\gev$. As
expected, the $e\mu \ell$ selections are more efficient than the $ee
\ell$ channels for scenarios with low mass difference between the
$\neut_1$ and the $\sse_R$. The $\mu\mu\ell$ channel may become
sensitive for models with $\mhalf \approx (370 - 380)\gev$, because
then the $\sstop_1$ decays dominantly via $\sstop_1\to\charge_1^+ b$,
and the decay of the chargino leads to an enhanced muon multiplicity,
\cf Table~\ref{Tab:SUSY2}. However, if the events are observed at the same
rate as in the current data, the $\mu\mu\ell$ channel will not play a
major r\^ole in testing $\sse_R$ LSP scenarios.

The D\O~analysis used in this paper, was optimized for associated
chargino and neutralino production within $R$-parity conserving
supersymmetry.  We point out, that larger regions of the selectron LSP
parameter space (compared to this paper) can be investigated by the
Tevatron collaborations if they optimize their cuts more towards our
scenarios. For example, a harder cut on the muon transverse momentum
will increase the signal to background ratio. In our models, the muons usually
stem from the decay of the (heavy) selectron LSP into two Standard
Model particles and thus have larger momenta. Similarly, a harder cut
on ${\not\!\!E_T}$ will help, since we have hard neutrinos stemming 
from the selectron or lightest stau decay and leading to a sizable amount of 
missing energy~\cite{Dreiner:2011wm}. We point out that an {\it upper}
cut on $H_T$, \textit{i.e.} the scalar sum of the transverse momenta of all jets,
should {\it not} be applied, because in large regions of the selectron
LSP parameter space, sparton pair production, which leads to hard jets
in the final state, occurs at a significant rate. 

We conclude this section by pointing out, that the D\O~analysis
is sensitive to an extended $\sse_R$ LSP parameter space with
future data. Under the (strong) assumption, that we can linearly
extrapolate the results given for an integrated luminosity of
$2.3~\mathrm{fb}^{-1}$ to a higher integrated luminosity, $\sse_R$ LSP
scenarios with $\mhalf\lesssim 450\gev$ may be probed with
$10~\mathrm{fb}^{-1}$.

\section{Summary and Conclusion}
\label{Sec:summary}

A right-handed selectron is a natural candidate for the LSP within the
$\text{B}_3$ mSUGRA model. If these or similar models are realized in
nature, they usually produce a strong signal of multi charged lepton
final states at hadron colliders like the Tevatron. On the one hand,
each selectron LSP decay produces one hard charged lepton and missing
energy.  On the other hand, the decays of heavier SUSY particles into
the selectron lead to additional charged leptons.

We have investigated the bounds on these models from the most recent
D\O~trilepton search (using an integrated luminosity of 2.3
fb$^{-1}$).  The non-observation of any events beyond the Standard
Model expectation puts stringent bounds on our models. We found that
scenarios with selectron LSP (squark) masses of up to $155\gev$
($880\gev$) are excluded. We also found that the selectron LSP region
consistent with the anomalous magnetic moment of the muon at $2\sigma$
(using spectral functions from $e^+e^-$ data) is ruled out by
the D\O~analysis. Thus, parameter regions outside B$_3$ mSUGRA
should also be considered, for example, non-universal scalar masses.

We then extrapolated the D\O~trilepton search to larger statistics,
\textit{i.e.}  assuming an integrated luminosity of 10 fb$^{-1}$. If
no excess over the Standard Model expectations is observed, the
Tevatron will be able to exclude selectron LSP models with a selectron
(squark) mass of up to $170\gev$ ($950\gev$).

\begin{acknowledgments}
We thank John Conley, Klaus Desch, Sebastian Fleischmann and Peter
Wienemann for helpful discussions.  S.G. thanks the Alexander von
Humboldt Foundation for financial support.  The work of S.G. was also
partly financed by the DOE grant DE-FG02-04ER41286. The work of
H.K.D. was supported by the BMBF ``Verbundprojekt HEP--Theorie'' under
the contract 05H09PDE and the Helmholtz Alliance ``Physics at the
Terascale''.
\end{acknowledgments}

\appendix
\section{Sparticle Masses and Branching Ratios of the Benchmark Models}
\label{App:benchmarks}

The benchmark scenarios SUSY1 and SUSY2 possess a light sparticle mass
spectrum. Therefore, the SUSY contribution to the anomalous magnetic
moment of the muon, $\delta a_\mu^\mathrm{SUSY}$, agrees within
$2\sigma$ with the discrepancy between the SM prediction and the
observation, \cf Eq.~(\ref{Eqn:amu}). In both benchmark scenarios,
SUSY1 and SUSY2, the $\sstau_1$ NLSP is nearly mass degenerate with
the $\sse_R$ LSP and exclusively undergoes the $R$-parity violating
decay $\sstau_1\to e\ssnumu$. The electrons from this decay usually
have a high momentum.

In Table~\ref{Tab:SUSY1}, we give the sparticle mass spectrum and
the dominant decay modes for SUSY1 ($\mzero=0\gev$, $\mhalf=400
\gev$, $\azero=-1250\gev$, $\tanb=5$, $\sgnmu = +$, $\lam_{231}|_
\mathrm{GUT}=0.045$). The $\sse_R$ LSP mass is about $139\gev$. Due to
the low $\mzero$ value, the mass difference between the $\neut_1$
next-to-NNLSP (NNNLSP) and the $\sse_R$ LSP is about $24\gev$ and thus
fairly large. The right-handed smuon, $\ssmu_R$, is the NNLSP and
undergoes three-body decays into the $\sse_R$ LSP and $\sstau_1$
NLSP. These decays are discussed in detail in
Ref.~\cite{Dreiner:2011wm} and usually yield a low-$p_T$ muon. The
lightest stop, $\sstop_1$, has a mass of $366\gev$ and decays
preferably into the $\neut_1$ and a $t$ quark. The first and second
generation squarks have masses around $820 - 860\gev$ and the gluino
mass is $934\gev$.

The sparticle mass spectrum and branching ratios of SUSY2 ($\mzero =
80\gev$, $\mhalf=375\gev$, $\azero=-1250\gev$, $\tanb=5$, $\sgnmu =
+$, $\lam_{231}|_\mathrm{GUT}=0.045$) are given in
Table~\ref{Tab:SUSY2}. This scenario lies near the $\neut_1$ LSP
region and thus the $\neut_1$ NNLSP, the $\sstau_1$ NLSP and the
$\sse_R$ LSP have nearly degenerate masses around
$152\gev$. We have a fairly light $\sstop_1$ with a mass of
$305\gev$. The $\sstop_1$ decay into the $\neut_1$ and a $t$ quark is
kinematically forbidden and $\sstop_1 \to \charge_1^\pm b$ is the only
decay mode. The squarks of the first and second generation have masses
around $ 780 - 820\gev$ and the mass of the gluino is $881\gev$.

\begin{table}
\scriptsize
\centering
\begin{tabular}{| lc | ll | ll |}
\hline
				&	mass [GeV]			&	channel				&	BR				&	channel				 &	BR		\\ 
\hline
$\sse_R^-$		&	$\mathbf{139.1}$		&	$\mu^-\nu_\tau$		&	$\mathbf{50\%}$	&	$\tau^- \nu_\mu$		&	$\mathbf{50\%}$	\\
\hline
$\sstau_1^-$		&	$139.6$				&	$e^- \bnumu$	&	$\mathbf{100\%}$			&		&		\\
\hline
$\ssmu_R^-$		&	$156.2$				&	$\sse_R^+ e^- \mu^-$	&	$30.2\%$ & $\sse_R^- e^+ \mu^-$		&	$25.1\%$	\\
				&						&	$\stau_1^+ \tau^- \mu^-$		&	$24.4\%$		& $\stau_1^- \tau^+ \mu^-$	& $20.3\%$		\\	
\hline
$\neut_1$			&	$163.3$				&	$\sse_R^- e^+$	&	$24.7\%$			&	$\sse_R^+ e^-$	&	$24.7\%$	\\
				&						&	$\stau_1^- \tau^+$	&	$22.9\%$			&	$\stau_1^+ \tau^-$	&	$22.9\%$	\\
				&						&	$\ssmu_R^- \mu^+$	&	$2.4\%$			&	$\ssmu_R^+ \mu^-$	&	$2.4\%$	\\
\hline
$\ssnutau$		&	$\mathbf{254.9}$		&	$\neut_1 \nu_\tau$		&	$63.9\%$			&	$W^+\sstau_1^-$		&	$24.1\%$		\\
				&						&	$e^- \mu^+$			&	$\mathbf{12.1\%}$		&						&			\\
\hline
$\ssnumu$		&	$\mathbf{258.1}$		&	$\neut_1 \nu_\mu$		&	$84.5\%$			&	$e^- \tau^+$			&	$\mathbf{15.5\%}$	\\
\hline
$\ssnue$			&	$262.9$				&	$\neut_1 \nu_e$		&	$100\%$			&						&			\\
\hline
$\ssmu_L^-$		&	$\mathbf{269.3}$		&	$\neut_1 \mu^-$		&	$84.2\%$			&	$e^- \bnutau$		&	$\mathbf{15.8\%}$	\\
\hline
$\sstau_2^-$		&	$\mathbf{269.6}$		&	$\neut_1 \tau^-$		&	$63.7\%$			&	$\Higgs \sstau_1^-$	&	$13.1\%$	\\
				&						&	$Z^0 \sstau_1^-$		&	$12.8\%$			&	$e^- \bnumu$		&	$\mathbf{10.5\%}$			\\
\hline
$\sse_L^-$		&	$273.9$				&	$\neut_1 e^-$		&	$100\%$			&			&		\\
\hline
$\neut_2$			&	$311.1$				&	$\ssbnutau \nutau$		&	$10.8 \%$			&	$\ssnutau \bnutau$		&	$10.8 \%$		\\
				&						&	$\ssbnumu \numu$		&	$9.7 \%$			&	$\ssnumu \bnumu$		&	$9.7 \%$		\\
				&						&	$\ssbnue\nue$			&	$8.1 \%$			&	$\ssnue \bnue$			&	$8.1 \%$		\\
				&						&	$\ssmu_L^- \mu^+$		&	$6.6\%$			&	$\ssmu_L^+ \mu^-$		&	$6.6\%$		\\
				&						&	$\sstau_2^- \tau^+$		&	$6.3\%$			&	$\sstau_2^+ \tau^-$		&	$6.3\%$		\\
				&						&	$\sse_L^- e^+$			&	$5.4\%$			&	$\sse_L^+ e^-$			&	$5.4\%$		\\		
				&						&	$\sstau_1^- \tau^+$		&	$2.7\%$			&	$\sstau_1^+ \tau^-$		&	$2.7\%$		\\
\hline
$\charge_1^-$		&	$311.2$				&	$\ssbnutau \tau^-$		&	$22.3\%$			&	$\ssbnumu \mu^-$		&	$20.0\%$		\\
				&						&	$\ssbnue e^-$			&	$16.9\%$			&	$\ssmu_L^- \bnumu$	&	$12.7\%$		\\
				&						&	$\sstau_2^- \bnutau$	&	$12.1\%$			&	$\sse_L^- \bnue$		&	$10.3\%$		\\
				&						&  	$\sstau_1^- \bnutau$	&	$5.0\%$			&						&				\\
\hline
$\sstop_1$		&	$365.8$				&	$\neut_1 t$			&	$69.1\%$			&	$\charge_1^+ b$		&	$30.9\%$		\\
\hline
$\ssbottom_1$		&	$706.3$				&	$W^- \sstop_1$		&	$78.5\%$			&	$\charge_1^- t$		&	$12.8\%$		\\
				&						&	$\neut_2 b$			&	$8.2\%$			&						&				\\
\hline
$\sstop_2$		&	$790.6$				&	$Z^0 \sstop_1$			&	$55.3\%$			&	$\Higgs \sstop_1$		&	$22.9\%$		\\
				&						&	$\charge_1^+ b$		&	$14.3\%$			&	$\neut_2 t$			&	$1.2\%$		\\
\hline
$\neut_3$			&		$819.8$			&	$\sstop_1 \bar{t}$		&	$26.5\%$			&	$\sstop_1^* t$			&	$26.5\%$		\\
				&						&	$\charge_1^- W^+$		&	$14.2\%$			&	$\charge_1^+ W^-$		&	$14.2\%$		\\
				&						&	$\neut_2 Z^0$			&	$12.6\%$			&	$\neut_1 Z^0$			&	$3.7\%$		\\
				&						&	$\neut_2 \Higgs$		&	$1.0\%$			&						&				\\
\hline
$\ssbottom_2$		&	$821.5$				&	$\neut_1 b$			&	$59.3\%$			&	$W^- \sstop_1$		&	$36.8\%$		\\
				&						&	$\charge_1^- t$		&	$2.0\%$			&	$\neut_2 b$			&	$1.2\%$		\\
\hline
$\ssdown_R\,(\ssstrange_R)$	&	$824.2$		&	$\neut_1 d (s) $		&	$100\%$			&						&				\\
\hline
$\ssup_R\,(\sscharm_R)$	&	$826.3$			&	$\neut_1 u (c) $		&	$100\%$			&						&				\\
\hline
$\charge_2^-$		&		$828.0$			&	$\sstop_1^* b$			&	$57.5\%$			&	$\neut_2 W^-$			&	$12.9\%$		\\
				&						&	$\charge_1^- Z^0$		&	$12.4\%$			&	$\charge_1^- \Higgs$	&	$11.6\%$		\\
				&						&	$\neut_1 W^-$			&	$3.3\%$			&						&				\\
\hline
$\neut_4$			&		$828.3$			&	$\sstop_1 \bar{t}$		&	$34.7\%$			&	$\sstop_1^* t$			&	$34.7\%$		\\
				&						&	$\charge_1^- W^+$		&	$9.0\%$			&	$\charge_1^+ W^-$		&	$9.0\%$		\\
				&						&	$\neut_2 \Higgs$		&	$7.5\%$			&	$\neut_1 \Higgs$		&	$2.2\%$		\\
\hline
$\ssup_L\,(\sscharm_L)$	&	$856.3$			&	$\charge_1^+ d (s) $	&	$65.9\%$			&	$\neut_2 u (c)$			&	$32.9\%$		\\
				&						&	$\neut_1 u (c)$			&	$1.2\%$			&						&				\\
\hline
$\ssdown_L\,(\ssstrange_L)$	&	$859.8$		&	$\charge_1^- u (c)$		&	$65.5\%$			&	$\neut_2 d (s)$			&	$32.8\%$		\\
						&				&	$\neut_1 d (s) $		&	$1.7\%$			&						&				\\
\hline
$\glu$			&		$933.8$			&	$\sstop_1 \bar{t}$		&	$23.4\%$			&	$\sstop_1^*  t $			&	$23.4\%$		\\
				&						&	$\ssbottom_1 \bar{b}$	&	$9.2\%$			&	$\ssbottom_1^* b $		&	$9.2\%$		\\
				&						&	$\ssbottom_2 \bar{b}$	&	$2.6\%$			&	$\ssbottom_2^* b $		&	$2.6\%$		\\
				&						&	$\ssdown_R \bar{d} (\ssstrange_R \bar{s})$	&	$2.5\%$	&	$\ssdown_R^* d (\ssstrange_R^* s ) $&	$2.5\%$	\\
				&						&	$\ssup_R \bar{u} (\sscharm_R \bar{c})$	&	$2.4\%$	&	$\ssup_R^* u (\sscharm_R^* c ) $&	$2.4\%$	\\
				&						&	$\ssup_L \bar{u} (\sscharm_L \bar{c})$	&	$1.3\%$	&	$\ssup_L^* u (\sscharm_L^* c ) $&	$1.3\%$	\\
				&						&	$\ssdown_L \bar{d} (\ssstrange_L \bar{s})$	&	$1.2\%$	&	$\ssdown_L^* d (\ssstrange_L^* s ) $&	$1.2\%$	\\
\hline
\end{tabular}
\caption{Branching ratios (BRs) and sparticle masses for the benchmark
scenario SUSY1. BRs smaller than $1\%$ are neglected.  $R$-parity
violating decays are shown in bold-face. Masses which are reduced by
more than 5 GeV (compared to the $R$-parity conserving spectrum) due
to $\lambda_{231}|_{\mathrm{GUT}} = 0.045$ are also shown in
bold-face.  }
\label{Tab:SUSY1}
\end{table}

\begin{table}
\scriptsize
\centering
\begin{tabular}{| lc | ll | ll |}
\hline
				&	mass [GeV]			&	channel				&	BR				&	channel				 &	BR		\\ 
\hline
$\sse_R^-$		&	$\mathbf{151.5}$		&	$\mu^-\nu_\tau$		&	$\mathbf{50\%}$	&	$\tau^- \nu_\mu$		&	$\mathbf{50\%}$	\\
\hline
$\sstau_1^-$		&	$151.6$				&	$e^- \bnumu$	&	$\mathbf{100\%}$			&		&		\\
\hline
$\neut_1$			&	$152.8$				&	$\sse_R^- e^+$	&	$50\%\%$			&	$\sse_R^+ e^-$	&	$50\%$	\\
\hline
$\ssmu_R^-$		&	$167.3$				&	$\neut_1 \mu^-$	&	$100\%$		&						&				\\ 
\hline
$\ssnutau$		&	$\mathbf{250.4}$		&	$\neut_1 \nu_\tau$		&	$75.7\%$			&	$e^- \mu^+$			&	$\mathbf{12.6\%}$		\\
				&						&	$W^+\sstau_1^-$		&	$11.7\%$		&						&			\\
\hline
$\ssnumu$		&	$\mathbf{253.7}$		&	$\neut_1 \nu_\mu$		&	$86.1\%$			&	$e^- \tau^+$			&	$\mathbf{13.9\%}$	\\
\hline
$\ssnue$			&	$258.5$				&	$\neut_1 \nu_e$		&	$100\%$			&						&			\\
\hline
$\ssmu_L^-$		&	$\mathbf{265.0}$		&	$\neut_1 \mu^-$		&	$85.5\%$			&	$e^- \bnutau$		&	$\mathbf{14.5\%}$	\\
\hline
$\sstau_2^-$		&	$\mathbf{265.6}$		&	$\neut_1 \tau^-$		&	$80.6\%$			&	$e^- \bnumu$		&	$\mathbf{11.6\%}$	\\
				&						&	$Z^0 \sstau_1^-$		&	$7.7\%$			&					&					\\
\hline
$\sse_L^-$		&	$269.6$				&	$\neut_1 e^-$		&	$100\%$			&			&		\\
\hline
$\neut_2$			&	$291.0$				&	$\ssbnutau \nutau$		&	$12.0 \%$			&	$\ssnutau \bnutau$		&	$12.0 \%$		\\
				&						&	$\ssbnumu \numu$		&	$10.3 \%$			&	$\ssnumu \bnumu$		&	$10.3 \%$		\\
				&						&	$\ssbnue\nue$			&	$7.9 \%$			&	$\ssnue \bnue$			&	$7.9 \%$		\\
				&						&	$\ssmu_L^- \mu^+$		&	$5.5\%$			&	$\ssmu_L^+ \mu^-$		&	$5.5\%$		\\
				&						&	$\sstau_2^- \tau^+$		&	$5.0\%$			&	$\sstau_2^+ \tau^-$		&	$5.0\%$		\\
				&						&	$\sstau_1^- \tau^+$		&	$4.7\%$			&	$\sstau_1^+ \tau^-$		&	$4.7\%$		\\
				&						&	$\sse_L^- e^+$			&	$3.8\%$			&	$\sse_L^+ e^-$			&	$3.8\%$		\\		
				&						&	$\neut_1 \Higgs$		&	$1.0\%$			&						&				\\

\hline
$\charge_1^-$		&	$291.0$				&	$\ssbnutau \tau^-$		&	$24.8\%$			&	$\ssbnumu \mu^-$		&	$21.3\%$		\\
				&						&	$\ssbnue e^-$			&	$16.4\%$			&	$\ssmu_L^- \bnumu$	&	$10.5\%$		\\
				&						&	$\sstau_2^- \bnutau$	&	$9.6\%$			&	$\sstau_1^- \bnutau$	&	$8.9\%$		\\
				&						&  	$\sse_L^- \bnue$		&	$7.2\%$			&	$\neut_1 W^-$			&	$1.0\%$		\\
\hline
$\sstop_1$		&	$304.9$				&	$\charge_1^+ b$		&	$100\%$			&						&				\\
\hline
$\ssbottom_1$		&	$661.6$				&	$W^- \sstop_1$		&	$80.9\%$			&	$\charge_1^- t$		&	$11.2\%$		\\
				&						&	$\neut_2 b$			&	$7.5\%$			&						&				\\
\hline
$\sstop_2$		&	$750.2$				&	$Z^0 \sstop_1$			&	$57.1\%$			&	$\Higgs \sstop_1$		&	$22.2\%$		\\
				&						&	$\charge_1^+ b$		&	$13.2\%$			&	$\neut_2 t$			&	$5.4\%$		\\
				&						&	$\neut_1 t$			&	$1.2\%$			&						&				\\
\hline
$\ssbottom_2$		&	$779.1$				&	$\neut_1 b$			&	$56.7\%$			&	$W^- \sstop_1$		&	$39.3\%$		\\
				&						&	$\charge_1^- t$		&	$1.9\%$			&	$\neut_2 b$			&	$1.2\%$		\\
\hline
$\ssdown_R\,(\ssstrange_R)$	&	$781.9$		&	$\neut_1 d (s) $		&	$100\%$			&						&				\\
\hline
$\ssup_R\,(\sscharm_R)$	&	$783.7$			&	$\neut_1 u (c) $		&	$100\%$			&						&				\\
\hline
$\neut_3$			&		$793.9$			&	$\sstop_1 \bar{t}$		&	$28.5\%$			&	$\sstop_1^* t$			&	$28.5\%$		\\
				&						&	$\charge_1^- W^+$		&	$13.0\%$			&	$\charge_1^+ W^-$		&	$13.0\%$		\\
				&						&	$\neut_2 Z^0$			&	$11.4\%$			&	$\neut_1 Z^0$			&	$3.3\%$		\\
				&						&	$\neut_2 \Higgs$		&	$1.0\%$			&						&				\\
\hline
$\charge_2^-$		&		$802.1$			&	$\sstop_1^* b$			&	$60.4\%$			&	$\neut_2 W^-$			&	$11.9\%$		\\
				&						&	$\charge_1^- Z^0$		&	$11.5\%$			&	$\charge_1^- \Higgs$	&	$10.7\%$		\\
				&						&	$\neut_1 W^-$			&	$3.0\%$			&						&				\\
\hline
$\neut_4$			&		$802.4$			&	$\sstop_1 \bar{t}$		&	$36.1\%$			&	$\sstop_1^* t$			&	$36.1\%$		\\
				&						&	$\charge_1^- W^+$		&	$8.2\%$			&	$\charge_1^+ W^-$		&	$8.2\%$		\\
				&						&	$\neut_2 \Higgs$		&	$6.7\%$			&	$\neut_1 \Higgs$		&	$2.0\%$		\\
\hline
$\ssup_L\,(\sscharm_L)$	&	$811.4$			&	$\charge_1^+ d (s) $	&	$66.0\%$			&	$\neut_2 u (c)$			&	$33.0\%$		\\
				&						&	$\neut_1 u (c)$			&	$1.0\%$			&						&				\\
\hline
$\ssdown_L\,(\ssstrange_L)$	&	$815.0$		&	$\charge_1^- u (c)$		&	$65.5\%$			&	$\neut_2 d (s)$			&	$32.8\%$		\\
						&				&	$\neut_1 d (s) $		&	$1.7\%$			&						&				\\
\hline
$\glu$			&		$881.0$			&	$\sstop_1 \bar{t}$		&	$24.7\%$			&	$\sstop_1^*  t $			&	$24.7\%$		\\
				&						&	$\ssbottom_1 \bar{b}$	&	$9.5\%$			&	$\ssbottom_1^* b $		&	$9.5\%$		\\
				&						&	$\ssbottom_2 \bar{b}$	&	$2.4\%$			&	$\ssbottom_2^* b $		&	$2.4\%$		\\
				&						&	$\ssdown_R \bar{d} (\ssstrange_R \bar{s})$	&	$2.3\%$	&	$\ssdown_R^* d (\ssstrange_R^* s ) $&	$2.3\%$	\\
				&						&	$\ssup_R \bar{u} (\sscharm_R \bar{c})$	&	$2.2\%$	&	$\ssup_R^* u (\sscharm_R^* c ) $&	$2.2\%$	\\
				&						&	$\ssup_L \bar{u} (\sscharm_L \bar{c})$	&	$1.2\%$	&	$\ssup_L^* u (\sscharm_L^* c ) $&	$1.2\%$	\\
				&						&	$\ssdown_L \bar{d} (\ssstrange_L \bar{s})$	&	$1.1\%$	&	$\ssdown_L^* d (\ssstrange_L^* s ) $&	$1.1\%$	\\
\hline
\end{tabular}
\caption{Same as Table~\ref{Tab:SUSY1}, but for the benchmark point SUSY2.} 
\label{Tab:SUSY2}
\end{table}

\end{document}